\documentclass[conference,a4paper]{IEEEtran}
\IEEEoverridecommandlockouts

\usepackage{cite}
\usepackage{amsmath,amssymb,amsfonts}
\usepackage{algorithmic}
\usepackage{graphicx}
\usepackage{textcomp}
\usepackage{xcolor}
\usepackage{booktabs}
\def\BibTeX{{\rm B\kern-.05em{\sc i\kern-.025em b}\kern-.08em
    T\kern-.1667em\lower.7ex\hbox{E}\kern-.125emX}}
\begin{document}
\setlength{\dbltextfloatsep}{6pt plus 1pt minus 2pt}

\title{Codec-Compatible Learned Compression for 3D Gaussian Splatting
\thanks{This paper is supported in part by National Natural Science Foundation of
China (62371290), National Key R\&D Program of China (2024YFB2907204),
the Fundamental Research Funds for the Central Universities of China,
and STCSM under Grant (22DZ2229005).
\textsuperscript{*}Corresponding authors: Yiling Xu
(e-mail: yl.xu@sjtu.edu.cn) and Qi Yang
(e-mail: qiyang@umkc.edu).}
}

\author{
\IEEEauthorblockN{
Feiyang Yu\textsuperscript{1},
Qi Yang\textsuperscript{2*},
Wenjie Huang\textsuperscript{1},
Yunfeng Guan\textsuperscript{1},
Yiling Xu\textsuperscript{1*},
and Zhu Li\textsuperscript{2}
}
\IEEEauthorblockA{
\textsuperscript{1}Shanghai Jiao Tong University, Shanghai, China,
\quad
\textsuperscript{2}University of Missouri-Kansas City, Kansas City, USA\\
\textsuperscript{1}\{yufy, huangwenjie2023, yfguan69, yl.xu\}@sjtu.edu.cn,
\quad
\textsuperscript{2}\{qiyang, lizhu\}@umkc.edu
}
}

\maketitle

\begin{abstract}
 Video-based 3D Gaussian Splatting (3DGS) compression methods provide a practical route by representing Gaussian attributes as 2D maps and coding them with standard video codecs. Among these maps, high-dimensional non-DC spherical harmonics (SH-AC) maps contain the largest number of channels and account for a dominant portion of the bitstream. However, their statistics differ substantially from those of natural images, and the nonlinear dependencies among SH-AC coefficients remain insufficiently exploited. To address these issues, we propose CCT-GS, a Codec-Compatible learned Transform framework for 3DGS compression. CCT-GS first learns a generalized SH-AC transform that captures cross-scene regularities and converts SH-AC maps into compact codec-friendly bottleneck maps. It further introduces LoRA-based scene adaptation to recover scene-specific appearance details with lightweight side information. Experiments on MPEG GSC sequences demonstrate that CCT-GS consistently outperforms representative video-based baselines, achieving an average BD-rate reduction of 51.30\% over the video-based anchor. The code is available at https://github.com/yfffffy/CCT-GS.
\end{abstract}


\section{Introduction}
3D Gaussian Splatting (3DGS)~\cite{kerbl3Dgaussians} has become a popular representation for novel view synthesis. By combining explicit Gaussian primitives with efficient differentiable rasterization, 3DGS enables high-fidelity rendering at interactive rates. However, a high-quality 3DGS scene can contain millions of primitives, each associated with multiple geometric and appearance attributes, resulting in substantial storage overhead \cite{yang2025hybridgs}. This makes efficient compression crucial for the practical storage and transmission of 3DGS content~\cite{xing2025vbench}.

Existing 3DGS compression methods can be broadly grouped into two categories aligned with the MPEG Gaussian Splat Coding (GSC) tracks: Alternate-track (A-3DGS) and Inria-track (I-3DGS)~\cite{mpegGSCRequirements2025}. A-3DGS methods adopt alternative compact representations, often built on structured anchors, compact learned features, feature-plane representations, or codebooks~\cite{scaffoldgs,hac,cat3dgs,contextgs,compact3dgs,c3dgs,codecgs,lightgaussian,compgs}. Many of them further employ dedicated entropy or context models for coding. These methods usually achieve strong rate-distortion (RD) performance, but their specialized representations, long coding time, and heavy resource requirement can limit compatibility and utilization of existing 3DGS pipelines. In contrast, I-3DGS methods operate on pretrained vanilla 3DGS models. Representative approaches include feed-forward compressors~\cite{fcgs,locomoco}, attribute-transform and hierarchical coding methods~\cite{adaptivevoxelization,mesongs,hgsc}, post-training pruning~\cite{pup3dgs}, and video-based methods~\cite{gscodec,lgscv,gsico,sog}. Among them, video-based methods are particularly attractive because they enable the reuse of mature video codecs. 

This work focuses on I-3DGS video-based methods, in which explicit Gaussian attributes are typically organized into 2D maps and compressed using standard video codecs. Among these maps, the non-DC spherical harmonics (SH-AC) coefficient maps contain the largest number of channels and represent a major source of coding overhead, making them a promising target for compression. However, SH-AC maps differ substantially from natural images in both spatial organization and channel statistics, limiting the coding efficiency achievable with standard video codecs. Existing methods have explored this problem from two complementary aspects: spatial map organization and channel-wise SH-AC compaction. SOGS~\cite{sog} and GSICO~\cite{gsico} reorganize Gaussian attributes into more spatially coherent 2D maps, while LGSCV~\cite{lgscv} and SHAC-PCA~\cite{shacpca} exploit linear inter-channel correlations among SH-AC coefficients through PCA-based dimensionality reduction. This direction has also attracted attention in recent MPEG GSC activities, where GSCodec Studio~\cite{mpegJEE62Anchor2026} serves as the video-based anchor and TMCV~\cite{TMCV2025} is developed as a candidate test model. Nevertheless, existing explicit attribute-map compression methods still rely largely on hand-designed map organization and linear transforms. A more effective approach is therefore needed to improve map compressibility while better matching the coding characteristics of standard codecs. Sandwiched Compression~\cite{sandwiched} provides a relevant precedent. Originally developed for image and video compression, it learns codec-friendly representations around a fixed standard codec. Its ability to repurpose conventional codecs for signals beyond their native design scope motivates us to extend this paradigm to SH-AC attribute-map compression. 
    
\begin{figure*}[t]
    \centering
    \includegraphics[width=1\textwidth]{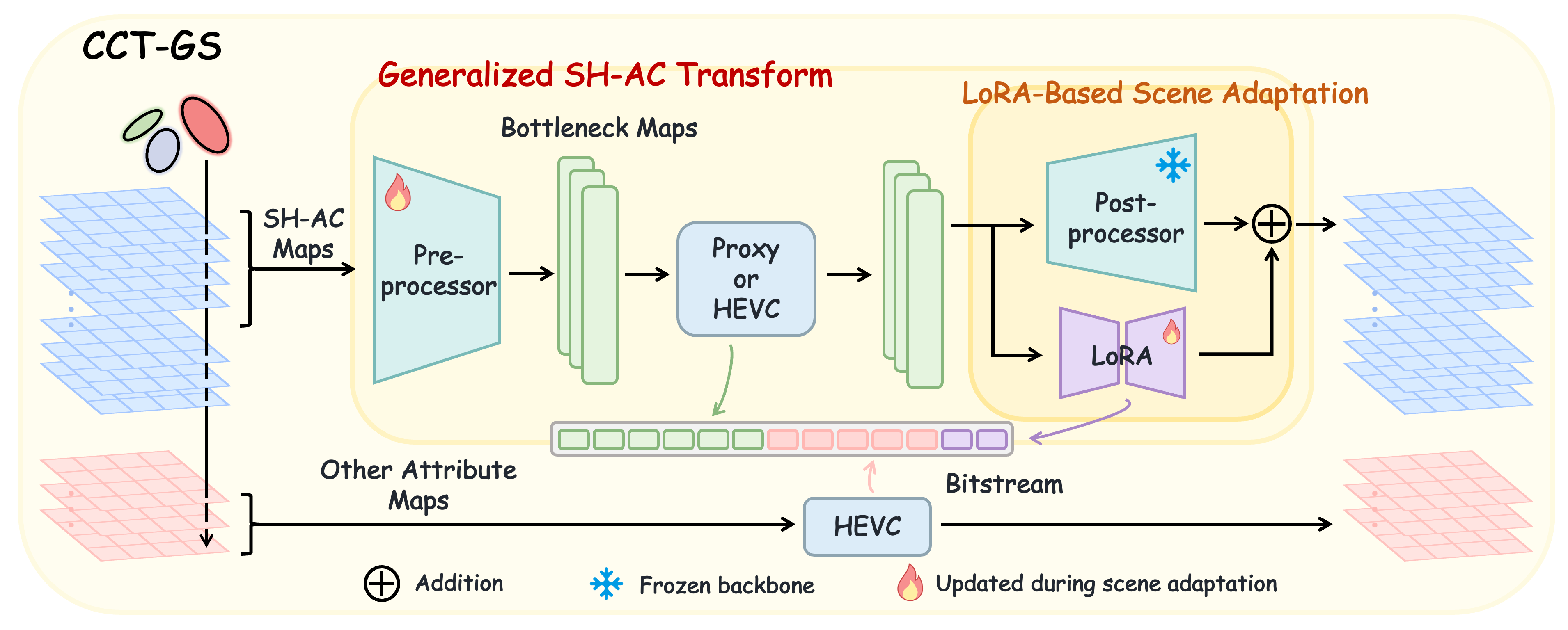}
    \caption{Overview of the proposed CCT-GS framework. A generalized transform converts SH-AC maps into codec-friendly bottleneck maps for HEVC coding, while LoRA-based scene adaptation refines their reconstruction. The remaining attribute maps are directly coded by HEVC.}
    \label{fig:pipeline}
\end{figure*}

In this paper, we propose CCT-GS, a \textbf{C}odec-\textbf{C}ompatible learned \textbf{T}ransform framework for 3D \textbf{G}aussian \textbf{S}platting compression. CCT-GS first learns a generalized SH-AC transform to convert 45-channel SH-AC attribute maps into compact and codec-friendly bottleneck maps. This transform models spatial and inter-channel regularities shared across scenes, substantially reducing the bitrate required for SH-AC coding while preserving compatibility with standard codecs. CCT-GS further incorporates a LoRA-based scene adaptation module to recover scene-specific appearance details that are not fully captured by the generalized model. Together, the two modules improve the rate-distortion performance of explicit SH-AC attribute-map coding. The main contributions are summarized as follows:

    \begin{itemize}
        \item We propose CCT-GS, a codec-compatible learned transform framework for explicit SH-AC attribute-map compression.
        \item We introduce LoRA-based scene adaptation to recover scene-specific appearance details.
        \item CCT-GS achieves superior RD performance, reducing the average BD-rate by 51.30\% over the video-based anchor.
    \end{itemize}

\begin{figure*}[!t]
    \centering
    \includegraphics[width=1\textwidth]{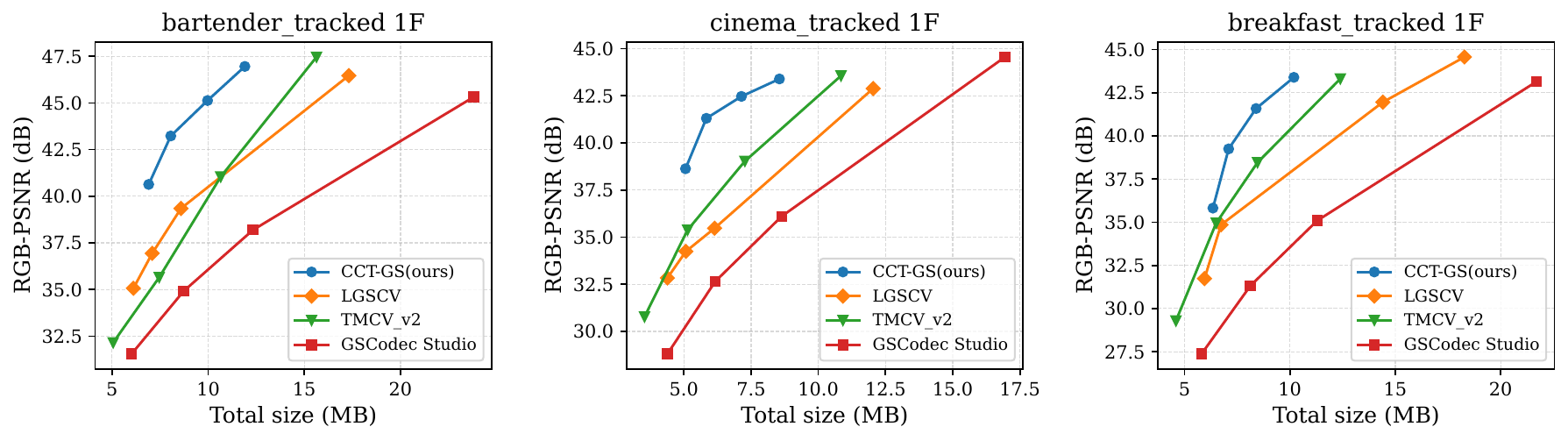}
    \caption{R-D performance on MPEG GSC sequences.}
    \label{fig:rd_curve}
\end{figure*}

\section{Proposed Method}

\subsection{Overview}
The overall pipeline of CCT-GS is illustrated in Fig.~\ref{fig:pipeline}. Given a pretrained vanilla 3DGS model $\mathcal{G}=\{\mathcal{G}_i\}_{i=1}^{N}$, each Gaussian $\mathcal{G}_i$ is parameterized by position $\boldsymbol{\mu}_i$, scale $\mathbf{s}_i$, rotation $\mathbf{q}_i$, opacity $\alpha_i$, and SH coefficients $\{\mathbf{c}^{\mathrm{dc}}_i,\mathbf{c}^{\mathrm{ac}}_i\}$, where $\mathbf{c}^{\mathrm{ac}}_i\in\mathbb{R}^{45}$. We organize these attributes into 2D maps using a two-stage Morton scan~\cite{lgscv}, obtaining the SH-AC maps $\mathbf{M}_{\mathrm{ac}}\in\mathbb{R}^{45\times H\times W}$ and the remaining attribute maps $\mathbf{M}_{\mathrm{oth}}$. In SH-AC branch, a pre-processor transforms $\mathbf{M}_{\mathrm{ac}}$ into compact bottleneck maps, which are encoded and decoded by HEVC. A post-processor then reconstructs the SH-AC maps $\hat{\mathbf{M}}_{\mathrm{ac}}$ from the decoded bottlenecks. Meanwhile, $\mathbf{M}_{\mathrm{oth}}$ is directly compressed and reconstructed using HEVC. For each target scene, CCT-GS performs scene-adaptive refinement with LoRA to recover scene-specific appearance details. Only the decoder-side LoRA parameters are transmitted as neural side information. The final bitstream is denoted by $\mathcal{B}=\{B_{\mathrm{ac}}, B_{\mathrm{oth}}, B_{\mathrm{LoRA}}\}$, corresponding to the SH-AC branch, the remaining attributes, and the compressed LoRA side information, respectively. Finally, the reconstructed attribute maps are merged and inversely reordered to yield the reconstructed 3DGS model. 

\subsection{Generalized SH-AC Transform}

To transform the 45-channel SH-AC maps $\mathbf{M}_{\mathrm{ac}}$ into compact bottleneck maps, we follow the learned pre- and post-processing architecture of Sandwiched Compression~\cite{sandwiched}, which employs a differentiable codec proxy for bottleneck reconstruction and bitrate estimation, thereby enabling joint optimization of the two processors to produce codec-friendly bottleneck maps. Each processor adopts a parallel structure with a $1\times1$ convolutional MLP branch and a U-Net branch. The $1\times1$ branch performs point-wise channel mixing and conversion, while the U-Net branch captures local spatial context in the attribute maps.
Let $f_{\theta}$ denote the pre-processor. Given $\mathbf{M}_{\mathrm{ac}}$, it generates the bottleneck maps $\mathbf{Y}\in\mathbb{R}^{15\times H\times W}$:
\begin{equation}
\mathbf{Y}=f_{\theta}(\mathbf{M}_{\mathrm{ac}}).
\end{equation}
The 15 bottleneck channels are grouped into five three-channel images in 4:4:4 format for HEVC coding. During training, the proxy $\mathcal{P}$ produces reconstructed bottleneck maps $\widetilde{\mathbf{Y}}$:
\begin{equation}
\widetilde{\mathbf{Y}}=\mathcal{P}(\mathbf{Y};\Delta),
\end{equation}
where $\Delta$ denotes the quantization step size used in the codec proxy. The reconstructed SH-AC maps $\hat{\mathbf{M}}_{\mathrm{ac}}$ are then obtained by the post-processor $g_{\phi}$:
\begin{equation}
\hat{\mathbf{M}}_{\mathrm{ac}}=g_{\phi}(\widetilde{\mathbf{Y}}).
\end{equation}
During actual compression, the proxy is replaced by the HEVC encoder and decoder. This transform is pretrained across multiple scenes and serves as the shared model for the subsequent scene-adaptation stage.

\subsection{LoRA-Based Scene Adaptation}

After obtaining the cross-scene pretrained model, we jointly adapt bottleneck generation and reconstruction for each target scene. The pre-processor is fine-tuned to generate scene-specific bottleneck maps, while the post-processor backbone remains frozen and only the inserted LoRA~\cite{lora} parameters are optimized to recover scene-specific information in the reconstructed SH-AC maps.

For each LoRA-enabled layer in the post-processor, let $\mathbf{W}$ denote its pretrained weight, with $d_{\mathrm{in}}$ and $d_{\mathrm{out}}$ denoting the input and output channel dimensions. LoRA parameterizes the adapted weight as
\begin{equation}
\mathbf{W}'
=
\mathbf{W}
+
\frac{\alpha}{r}\mathbf{B}\mathbf{A},
\end{equation}
where $\mathbf{A}$ and $\mathbf{B}$ are trainable low-rank factors, $r\ll\min(d_{\mathrm{in}},d_{\mathrm{out}})$ is the LoRA rank, and $\alpha$ is a scaling hyperparameter. During scene adaptation, $\mathbf{W}$ remains frozen, and only the LoRA factors $\mathbf{A}$ and $\mathbf{B}$ are updated for the post-processor. Since the post-processor is instantiated as a convolutional network, this low-rank update is implemented as a residual branch consisting of a $1\times1$ down-projection convolution followed by an up-projection convolution with the same kernel size as the original convolution. LoRA is inserted into the convolutional layers in the post-processor, and we denote all scene-specific LoRA parameters for scene $s$ by $\boldsymbol{\psi}_s$. The adapted pre-processor and $\boldsymbol{\psi}_s$ are jointly optimized using the scene-adaptation objective described in the following subsection. At the decoder, the shared post-processor backbone is augmented with $\boldsymbol{\psi}_s$ to reconstruct the SH-AC maps as $\widehat{\mathbf{M}}_{\mathrm{ac}}^{(s)}
=
g_{\phi,\boldsymbol{\psi}_s}(\widetilde{\mathbf{Y}}^{(s)})$. Since the adapted pre-processor is used only to generate the coded bottleneck maps at the encoder side, its parameters need not be transmitted. Therefore, only $\boldsymbol{\psi}_s$ is conveyed as model side information, contributing $B_{\mathrm{LoRA}}$ to the total bitstream.

\subsection{Loss Function}

We optimize CCT-GS with a rate-distortion objective. The codec proxy estimates the bottleneck coding rate $\widehat{R}_{\mathrm{ac}}$, and the objective is formulated as
\begin{equation}
\mathcal{L}_{\mathrm{RD}}
=
\mathcal{D}
+
\lambda_{\mathrm{R}}\widehat{R}_{\mathrm{ac}},
\end{equation}
where $\lambda_{\mathrm{R}}$ is a hyperparameter balancing the distortion and rate terms, and $\mathcal{D}$ denotes the distortion loss.

For map-domain supervision, we define the distortion between the reconstructed and original SH-AC maps as
\begin{equation}
\mathcal{L}_{\mathrm{map}}
=
\frac{1}{45HW}
\left\|
\widehat{\mathbf{M}}_{\mathrm{ac}}
-
\mathbf{M}_{\mathrm{ac}}
\right\|_2^2.
\end{equation}
Since map-domain distortion does not necessarily reflect the perceptual impact of SH-AC reconstruction errors, we further employ a render-aware photometric loss to directly supervise the rendered images. Following the original 3DGS training objective~\cite{kerbl3Dgaussians}, the render-aware loss combines an $L_1$ term and a structural similarity term:
\begin{equation}
\begin{aligned}
\mathcal{L}_{\mathrm{render}}
=&\,
(1-\lambda_{\mathrm{ssim}})
\left\|
\widehat{\mathbf{I}}-\mathbf{I}
\right\|_1 \\
&+
\lambda_{\mathrm{ssim}}
\left[
1-\mathrm{SSIM}
\left(
\widehat{\mathbf{I}},\mathbf{I}
\right)
\right].
\end{aligned}
\end{equation}
where $\widehat{\mathbf{I}}$ and $\mathbf{I}$ denote the rendered and ground-truth images, respectively, and $\lambda_{\mathrm{ssim}}$ is a hyperparameter balancing the two photometric terms. In practice, the loss is averaged over sampled training views.

During cross-scene pretraining, $\mathcal{D}$ is first set to $\mathcal{L}_{\mathrm{map}}$ to stabilize the learned transform, and then replaced by $\mathcal{L}_{\mathrm{render}}$ for render-aware refinement. During scene adaptation, $\mathcal{D}$ is set to $\mathcal{L}_{\mathrm{render}}$ to optimize the adapted pre-processor and scene-specific LoRA parameters.

\section{Experimental results}
\subsection{Experimental Setup}

We evaluate our method following the MPEG GSC CTC~\cite{CTC2026} on the full Forward-facing class in the 1F setting. Objective quality is evaluated using the mpeg-gsc-metric tool according to the CTC protocol. For \emph{breakfast}-prefixed sequences, views 6 and 8 are used for testing; for the remaining forward-facing sequences, views 9 and 11 are used. The non-test views are used for scene adaptation.

The generalized transform is trained on 75 scenes from the DL3DV-GS dataset~\cite{fcgs,dl3dv}. We set $\lambda_{\mathrm{R}}=0.5$ for the rate-distortion objective. In cross-scene pretraining, we optimize $\mathcal{L}_{\mathrm{RD}}$ with $\mathcal{D}=\mathcal{L}_{\mathrm{map}}$ for the first 600 epochs and with $\mathcal{D}=\mathcal{L}_{\mathrm{render}}$ for the next 200 epochs. During scene adaptation, we optimize $\mathcal{L}_{\mathrm{RD}}$ with $\mathcal{D}=\mathcal{L}_{\mathrm{render}}$ on the training views. LoRA uses rank $r=8$ and scaling parameter $\alpha=8$. All neural training is conducted on a single NVIDIA RTX 4090 GPU.

We use the HEVC reference software HM 18.0~\cite{hm18} for video coding and DeepCABAC~\cite{deepcabac} for entropy coding of the LoRA side information.

\subsection{Compression Performance}

We compare CCT-GS with representative video-based 3DGS compression baselines, including LGSCV~\cite{lgscv}, GSCodec Studio~\cite{mpegJEE62Anchor2026}, and TMCV-v2~\cite{TMCV2025}. For CCT-GS, we use the MPEG proposal configuration adopted in LGSCV~\cite{lgscv} for explicit attribute-map coding, including the rate points, bit depths, and map packing settings. CCT-GS keeps the same settings for non-SH-AC attributes and encodes the learned bottleneck maps using the corresponding SH-AC configuration. For LGSCV, we follow its reported configuration. GSCodec Studio and TMCV-v2 are evaluated using their official configurations~\cite{CTC2026,TMCV2025}. We report RGB-PSNR versus compressed size in MB, where the size includes all coded files, metadata, and side information required for decoding.

Fig.~\ref{fig:rd_curve} presents representative R-D curves on three tracked sequences. CCT-GS consistently achieves better R-D performance than all baselines, with the curves shifted toward the upper-left region. Compared with LGSCV, the clear improvement shows that the learned SH-AC bottleneck transform is more effective for SH-AC compression than PCA-based linear dimensionality reduction. Compared with GSCodec Studio and TMCV-v2, CCT-GS also achieves lower bitrate at comparable reconstruction quality, indicating that the generalized transform and scene-specific LoRA adaptation can better exploit the redundancy of SH-AC attributes while remaining compatible with standard video coding.

\begin{table}[!htbp]
\centering
\caption{BD-rate (\%) with GSCodec Studio as anchor on MPEG GSC sequences.}
\label{tab:bd_rate_gscodec}
\setlength{\tabcolsep}{2.0pt}
\renewcommand{\arraystretch}{1.05}
\begin{tabular*}{\columnwidth}{@{\extracolsep{\fill}}lrrrrrrr@{}}
\toprule
Method & BT & BS & CT & CS & BFT & Avg. & Overall \\
\midrule
LGSCV & -37.24 & -28.00 & -14.71 & -29.69 & -34.52 & -28.83 & -27.66 \\
TMCV-v2     & -30.00 & -32.13 & -35.06 & -33.71 & -41.31 & -34.44 & -34.11 \\
CCT-GS      & \textbf{-58.75} & \textbf{-52.75} & \textbf{-55.81} & \textbf{-49.18} & \textbf{-54.45} & \textbf{-54.19} & \textbf{-51.30} \\
\midrule
Method & BFS & BFU & Makeup & Musical & Trio & Avg. & Overall \\
\midrule
LGSCV & -43.38 & -38.86 &  -3.84 & -39.57 &  -6.78 & -26.49 & -27.66 \\
TMCV-v2     & -41.60 & -30.88 & -30.41 & -32.19 & -33.79 & -33.77 & -34.11 \\
CCT-GS      & \textbf{-53.03} & \textbf{-49.24} & \textbf{-44.29} & \textbf{-57.03} & \textbf{-38.50} & \textbf{-48.42} & \textbf{-51.30} \\
\bottomrule
\multicolumn{8}{@{}p{\columnwidth}@{}}{\scriptsize
BT: bartender tracked; BS: bartender semitracked; CT: cinema tracked; CS: cinema semitracked; BFT/BFS/BFU: breakfast tracked/semitracked/untracked.
}
\end{tabular*}
\end{table}

Table~\ref{tab:bd_rate_gscodec} reports BD-rate results over the full Forward-facing class using GSCodec Studio as the anchor. CCT-GS achieves consistent bitrate savings on all tested sequences and obtains the best average BD-rate reduction of 51.30\% relative to the anchor. The gains are observed across tracked, semi-tracked, and untracked sequences, demonstrating robust R-D improvements across different forward-facing 3DGS contents.

\subsection{Ablation Study}
\begin{figure}[!htbp]
    \centering
    \includegraphics[width=1\columnwidth]{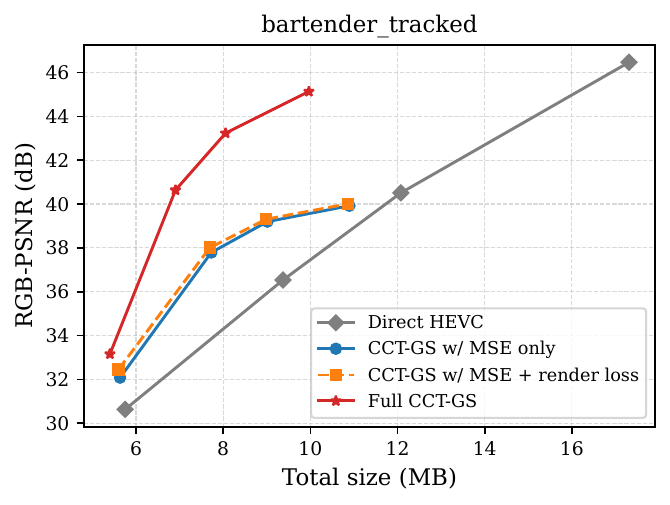}
    \caption{Ablation study on the contribution of each component.}
    \label{fig:ablation}
\end{figure}

We conduct an ablation study on the \emph{bartender\_tracked} sequence to validate the contribution of each component. Direct HEVC denotes directly coding the SH-AC maps with HEVC. As shown in Fig.~\ref{fig:ablation}, CCT-GS with MSE-only training (i.e., only generalized transform without LoRA-based adaptation)  achieves an 11.40\% BD-rate reduction over Direct HEVC, confirming the effectiveness of the learned SH-AC bottleneck transform. Introducing the rendering loss further brings a 3.13\% BD-rate reduction over the MSE-only model, indicating that render-aware optimization better aligns the reconstructed SH-AC attributes with the final view quality. With LoRA-based scene adaptation, the full CCT-GS obtains an additional 22.72\% BD-rate reduction over the generalized SH-AC transform without LoRA, and achieves a total BD-rate reduction of 37.75\% over Direct HEVC. These results show that the generalized SH-AC transform captures cross-scene regularities of SH-AC attributes, while LoRA-based scene adaptation further learns scene-specific appearance details to improve R-D efficiency.

\section{Conclusion}
In this paper, we proposed CCT-GS, a codec-compatible learned compression framework for explicit 3DGS attribute maps, focusing on the high-dimensional SH-AC attributes. By learning cross-scene regularities of SH-AC maps and further capturing scene-specific appearance details, CCT-GS improves R-D efficiency while retaining compatibility with standard video coding. Experiments under the MPEG GSC CTC demonstrate consistent gains over representative video-based baselines, achieving an average BD-rate reduction of 51.30\% over the video-based anchor. In future work, we plan to extend the learned transform to other Gaussian attributes and reduce the complexity of scene adaptation for practical 3DGS storage and transmission.



\bibliographystyle{IEEEtran}
\bibliography{reference}

\end{document}